\documentclass[superscriptaddress,pra,twocolumn,aps,floatfix,10pt]{revtex4-2}
\usepackage{amsmath}
\usepackage{amsfonts}
\usepackage{amssymb}
\usepackage{graphicx,graphics}
\usepackage{bm,bbm}
\usepackage{color}
\usepackage[colorlinks=true,linkcolor=blue,urlcolor=blue,citecolor=blue]
  {hyperref}
\DeclareGraphicsExtensions {.pdf,.png,.jpg} % search in this order!
\usepackage[justification=justified,
format=plain]{caption}
\usepackage{subfig}
\newcommand{\la}{\langle}
\newcommand{\ra}{\rangle}

\newcommand{\ket}[1]{|#1\rangle}
\newcommand{\bra}[1]{\langle #1|}
\newcommand{\One}{\mathbf{1}}

\newlength{\subcolumnwidth}

\newcommand{\nextsubcolumn}[1][]{%
	\cr\noalign{\hfill}
	\if\relax\detokenize{#1}\relax\else\hsize=#1\setlength{\subcolumnwidth}{\hsize}\fi
}

\begin{document}
\title{Normal quantum channels and Markovian correlated two-qubit quantum 
  errors}

\author{Alejandro Contreras Reynoso}
\email{alejandro.contreras8829@alumnos.udg.mx}
\affiliation{Departamento de F\'\i sica, Universidad de Guadalajara,
   Guadalajara, Jalisco, M\' exico.}
\author{Thomas Gorin}
\email{thomas.g@academicos.udg.mx}
\affiliation{Departamento de F\'\i sica, Universidad de Guadalajara,
   Guadalajara, Jalisco, M\' exico.}

\begin{abstract}
We study general ``normally'' distributed random unitary transformations. These 
distributions can be defined in terms of a diffusive random walk in the 
respective group manifold, formally underpinned by the concept of infinite 
divisibility. On the one hand, a normal distribution induces a unital quantum 
channel. On the other hand, the diffusive random walk defines a unital quantum 
process, which can be generated by a Lindblad master equation. In the single 
qubit case, we show that it is possible to find different distributions which 
induce the same quantum channel. 

In the case of two qubits, the normal quantum channels, i.e. quantum channels
induced by normal distributions in ${\rm SU}(2)\otimes{\rm SU}(2)$ provide an
appropriate framework for modeling correlated quantum errors. In contrast to
correlated Pauli errors, for instance, they conserve their Markovianity, and 
they lead to very different results in error correcting codes or entanglement
distillation. We expect our work to find applications in the tomography and 
modeling of one- and two-qubit errors in current quantum computer platforms, 
but also in the distillation of Bell pairs across imperfect communication 
channels, where it is conceivable that subsequently transmitted qubits are 
subject to correlated errors.
\end{abstract}

\maketitle

\section{\label{I} Introduction}

The concept of Gaussian or normally distributed errors is ubiquitous in science
and technology, where students are introduced to this concept, in undergraduate 
courses~\cite{ross_2014}. The most basic techniques of error estimation in 
experiments or observations are based on the assumption of a normally 
distributed error. Surprisingly a similarly flexible and general error model
is still missing in the area of quantum communication and quantum computing. 
Most commonly, error correction protocols, communication protocols, etc., are 
validated against uncorrelated single qubit Pauli errors~\cite{NieChu00,KLV00}. 

In the case of uncorrelated single-qubit errors, normal quantum channels (i.e. 
quantum channels induced by normal distributions) are essentially equivalent to 
Pauli errors. However, in the case of correlated two-qubit errors this is no 
longer the case. Correlated qubit errors may arise in quantum communication 
protocols, where several qubits must be sent in order to perform error 
correction or entanglement distillation. They also occur in current quantum 
computing platforms, for instance in the implementation of control-not 
gates~\cite{two-qubit_spectroscopy_noise,preskill2012sufficient}. Typically,
these two-qubit errors largely dominate the accumulated error in the 
realization of a given quantum algorithm \cite{yan2018tunable,Lao_2022}. 
In spite of this, correlated errors are only rarely considered in the 
literature, and if so, rather particular models are 
used~\cite{CorrelatedCaruso,CorrelatedMachiavelo}.

In this work we aim at constructing a model for quantum errors, following the 
basic principles of the classical normal distributions. In the case of a single 
qubit, the unitary transformations are associated to the compact Lie group 
${\rm SU}(2)$. Thus, we need to define normal distributions on this group 
manifold. To the best of our knowledge, the closest works in this respect are 
about normal distributions in ${\rm SO}(3)$ with applications in estimating
uncertainties in the orientation of rigid bodies~\cite{Par64,NikSav97}. We then 
move on to normal distributions in ${\rm SU}(2)\otimes{\rm SU}(2)$, which 
allow us to describe correlated errors affecting two qubits. 

The paper is organized as follows: After this introduction, in Sec.~\ref{aN}, 
we generalize the central results of~\cite{Par64,NikSav97} to arbitrary compact 
Lie groups. We find a parametrization for the family of normal distributions, 
together with an algebraic expression for their moments. This expression allows 
us to compute the quantum channel induced by any particular normal 
distribution. In Sec.~\ref{S}, we consider the single qubit case. We find a 
simple correspondence between the normal distributions centered on the 
identity, and general Pauli channels. Furthermore, we find that for any 
Lindblad-divisible quantum channel~\cite{DaZiPi19}, there are one or more 
normal distributions which generate this channel. In Sec.~\ref{T}, we study 
normal distributions in the group ${\rm SU}(2)\otimes{\rm SU}(2)$, to describe 
random errors in two-qubit operations. By construction, these distributions 
induce channels which can contain classical correlations, only (this is the 
main practical difference to normal distributions in ${\rm SU}(4)$). In 
Sec.~\ref{sA}, we illustrate the effect of correlated normal errors in a simple 
protocol for entanglement distillation. We conclude our paper with 
Sec.~\ref{sC}.

\section{\label{aN} Normal distributions on compact Lie groups} 

In probability theory, a probability distribution is infinitely divisible if it
can be expressed as the probability distribution of the sum of an arbitrary 
number of independent and identically distributed (i.i.d.) random 
variables~\cite{KenIti99,Yas00,dominguez_divisibility}.

In order to apply this concept to transformation groups, the sum must be 
replaced by the group operation. In addition, we require a unique invariant 
probability measure on the group, and we use some tools of harmonic analysis 
and there in particular the Peter-Weyl theorem. For these reasons, we restrict
ourselves to the case of compact Lie groups~\cite{Chirikjian12}.

\subsection{Divisibility}

Let $G$ be a compact Lie group, $g\in G$ a group element, and $dg$ the 
left-invariant normalized Haar measure~\cite{Haar33}. Then the probability 
density $f(g)$ is two-times divisible if there exists a probability density 
$h(g)$, such that
\begin{align}
  f(g) &= \int dg_1\int dg_2\; h(g_1)\; h(g_2)\; \delta(g - g_1 g_2)\notag\\
       &= \int d g_1\; h(g_1)\; h(g_1^{-1}\, g)
\quad\Leftrightarrow\quad f = h \star h \; .
\end{align}
Similarly, $f(g)$ is $n$-times divisible, if there exists a $h(g)$, such that 
\begin{equation}
  f = g^{\star n} 
    = \underbrace{h \star h \star \ldots \star h}_{n\; \text{terms}} \; .
\label{aN:nfoldConv}\end{equation}
This means that the sequence of $n$ i.i.d. random transformations 
$k_1, \ldots k_n$ with probability density $h(k_j)$ generate a random 
transformation $g = k_1 k_2 \ldots k_n$ with probability density $f(g)$. This 
sequence of transformations form a Markovian process which can be understood as 
a random walk in the group manifold, where each step is one of the $n$ 
transformations $k_j$.

\subsection{\label{aNH} Harmonic analysis}

Let us now consider the Hilbert space $\mathcal{L}^2(G)$ of square integrable 
functions $f: G \to \mathbb{C}$ on the group $G$. 
%We consider the compact topological group $G$ whose elements are $g$. 
Let $D^{(s)}(g)$ be an irreducible matrix representation of finite dimension
(in what follows, the index $s$ will be used to identify the elements of a 
whole sequence of representations). Then, we define the Fourier integral with 
respect to the kernel $D^{(s)}(g)$ as
\begin{equation}
\mathcal{F}_s[f] = \int dg\; f(g)\; D^{(s)}(g)\; ,
\label{aNH:DefFouInt}\end{equation}
where the integration on the R.H.S. must be applied to each matrix element of 
$D^{(s)}(g)$, such that $\mathcal{F}_s[f]$ is a matrix of coefficients of the 
same dimension as $D^{(s)}(g)$.

This Fourier integral fulfills a convolution theorem, just as ordinary Fourier
integrals do:
\begin{equation}
f = h_1 \star h_2 \quad\Leftrightarrow\quad
\mathcal{F}_s[f] = \mathcal{F}_s[h_1]\, \mathcal{F}_s[h_2]\; .
\label{fourier2}\end{equation}
Then, since $G$ is a compact Lie group, the Peter-Weyl 
theorem~\cite{peter_weyl_1927} asserts that (i) there exists an orthogonal sum 
of irreducible finite-dimensional unitary representations which we denote by 
$\oplus_{s=1}^\infty D^{(s)}(g)$, and (ii) that the matrix elements of all
$D^{(s)}(g)$ form an orthonormal basis in $\mathcal{L}^2 \left(G\right)$. 
Therefore, we can decompose the probability density $f$ in terms of Fourier 
matrix-coefficients
\begin{align}
f(g) &= \sum_s \sum_{m,m'} C^{(s)}_{m,m'}\; D^{(s)}(g)_{m,m'}^* \notag\\
 &= \sum_s {\rm tr}\big ( C^{(s)}\, D^{(s)}(g)^\dagger\big )\; , \qquad
C^{(s)} = \mathcal{F}_s[f]\; ,
\label{aNH:PWONB}\end{align}
and apply the convolution theorem to relate the Fourier coefficients of $f$ to 
the Fourier coefficients of $h$, when $f= h^{\star n}$.

\subsection{Normal distributions on \boldmath $G$}

As a compact Lie group, $G$ has a finite number of generators 
$\hat L_1, \ldots, \hat L_r$, which may be assumed to form an orthonormal basis
in the associated real Lie algebra \cite{CompactAlgebraReal}. Then, for any irreducible representation 
$D^{(s)}(g)$ we have corresponding representations of the generators, 
$L^{(s)}_1, \ldots, L^{(s)}_r$, which we collect into the vector
$\vec L^{(s)} = ( L^{(s)}_1, \ldots, L^{(s)}_r)$.

This allows to assign to any group element $g$ (and its representations 
$D^{(s)}(g)$) one (or several) points $\vec n$ in a $r$-dimensional real vector 
space, such that
\begin{align}
g &= \exp\Big ( -i\, \sum_{\nu = 1}^r n_\nu\, \hat L_\nu\Big )\notag\\
\Leftrightarrow\quad
D^{(s)}(g) &= \exp\big ( -i\, \vec n \cdot \vec L^{(s)}\big )\; .
\label{aN:expomap}\end{align}
Following Ref.~\cite{NikSav97}, we wish to define the family of normal 
distributions, as those probability densities $f\in\mathcal{L}^2(g)$, for which 
there exist a sequence of probability densities 
$\mathbb{N} \ni n \to h_n\in \mathcal{L}^2(g)$ such that 
\begin{equation}
f = \lim_{n\to\infty} h_n^{\star n}\; .
\label{aN:DefNormDist}\end{equation}
Let us consider some $n \gg 1$ fixed. Then, $f$ is the probability density for
a random transformation generated by a sequence of transformations, 
$k_1\ldots k_n$. Because the manifold is compact all moments of $f$ exist and
are finite. Therefore, the individual transformations $k_j$ must be close to 
the identity. In fact, as $n\to\infty$, they must move closer and closer to it.
This eventually allows us to replace the Haar measure $dg$ by the flat measure
in the in the tangent space of the group manifold. Therefore,
\begin{equation}
\int dg\; h_n(g)\; D^{(s)}(g) \approx \int d^r\vec n\; h_n(g)\;
   D^{(s)}(g)\; , 
\label{G:flatav}\end{equation}
where $g\equiv g(\vec n)$ is given by the exponential map introduced in 
Eq.~(\ref{aN:expomap}). It also allows us to use
the Taylor expansion of the exponential map. Up to second order this gives,
\begin{equation}
D^{(s)}(g) \approx \One - i\, \vec n \cdot \vec L^{(s)} - \frac{1}{2}\;
   \big ( \vec n \cdot \vec L^{(s)}\big )^2\; .
\end{equation}
Inserting this into Eq.~(\ref{G:flatav}), we obtain
\begin{align}
&\mathcal{F}_s(h_n) \approx \One -i\, \la\vec n\ra \cdot \vec L^{(s)} 
   - \frac{1}{2}\sum_{j,k} \langle n_j n_k\ra \, L^{(s)}_j\, L^{(s)}_k \\
&\quad = \One -i\, \la\vec n\ra \cdot \vec L^{(s)}
   - \frac{1}{2}\sum_{j,k} \big ( C_{jk}  + \la n_j\ra \la n_k\ra\, \big )\, 
   L^{(s)}_j\, L^{(s)}_k\; , \notag
\end{align}
where we use $\la\, \cdots\, \ra$ to denote the average over the distribution
$h_n$ and $C_{jk}$ are the matrix elements of the covariance matrix $\bm C$.
In order to make sure that the limit in Eq.~(\ref{aN:DefNormDist}) exists, we 
require that for large $n$, both, $\la\vec n\ra$ and $\bm C$ scale as $n^{-1}$. 
Then, we may neglect the product $\la n_j\ra \la n_k\ra$ and 
\begin{equation}
\mathcal{F}_s(h_n) \approx \One -i\, \la\vec n\ra \cdot \vec L^{(s)}
   - \frac{1}{2}\, \vec L^{(s)}{}^T\cdot \bm C\cdot \vec L^{(s)}\; .
\end{equation}
This finally gives
\begin{align}
\mathcal{F}_s[f] &= \lim_{n\to\infty} \mathcal{F}_s[h_n]^n 
   = e^{- \mathcal{M}_s}\notag\\
\mathcal{M}_s &= \frac{1}{2}\, 
     \vec L^{(s)}{}^T\cdot \bm A\cdot \vec L^{(s)} 
      + i\, \vec b \cdot \vec L^{(s)}\; ,
\label{aN:parametrization}\end{align}
where we call $\bm A = n\, \bm C$ the diffusion matrix and 
$\vec b = n\, \la \vec n\ra$ the drift vector. The diffusion matrix describes 
the width of the distribution and the drift vector its center. The 
expression in Eq.~(\ref{aN:parametrization}) provides a parametrization of the 
normal distributions in $G$, completely analogous to the multi-dimensional flat 
case, with a non-negative covariance matrix $\bm A$ and a displaced center 
$\vec b$. However, four remarks are in order:
\paragraph{} As we will see below, the interpretation of $\bm A$ and
$\vec b$ can be very complicated for non-abelian groups. For instance, taking
the average of a group element over the distribution, will normally not yield 
$\vec b$ as the center of the distribution. In fact the average of a group
element will typically no longer be an element of the group.
\paragraph{} It is clear that the parametrization in 
Eq.~(\ref{aN:parametrization}) leads to a distribution which is infinitely 
divisible and hence a normal distribution according to our definition. We speak 
of a ``diffusive random walk'' in this case. However, 
it is not clear whether all infinitely divisible distributions allow for such
a parametrization. In the $SO(3)$ case this has been shown in 
Refs.~\cite{Par64,NikSav97} but for the general cases outlined here, such a 
proof is yet missing, but beyond the scope of the present work.  On
the real line for instance there are infinitely divisible distributions which 
are not Gaussian \cite{ComunicacionFrancois}.
\paragraph{} In this work, we define normal distributions as those which allow 
for a parametrization as in Eq.~(\ref{aN:parametrization}). In this sense, the 
normal distributions are the continuation of the Gaussian distributions in the 
tangent space of the group manifold to the group manifold itself. This is then 
closely related to definitions of normal distributions as solutions of the 
diffusion equation (in this case, on the group manifold)~\cite{Par64,NikSav97}.
\paragraph{} Instead of the normal distribution $f$, we may choose the 
diffusive random walk with distribution $h_n$ as the starting point of our 
considerations. In this way, we may arrive at a Markovian quantum process,
which can be described by a Lindblad master equation in the continuum limit.
In the following Sec.~\ref{MEG}, this will be discussed in more detail.

\section{\label{MEG} General normal quantum channels}

For a compact Lie group $G$ with a unitary irreducible matrix representation
$U(g)=D^{(s)}(g)$ and normal distribution $f$ parameterized by a diffusion
matrix $\bm A$ and a drift vector $\vec b$, we define a general normal quantum 
channel as
\begin{equation}
\Lambda(\bm A,\vec b\, ) \quad : \quad \varrho \to 
   \int dg\; f(g)\; U(g)\, \varrho\, U(g)^\dagger\; ,
\label{generalnormalmap}\end{equation}
where $\varrho$ is a density matrix of finite dimension, related to the 
parameter $s$. A case of particular interest will be considered below, in 
Sec.~\ref{S}, where $U(g)$ is interpreted as a random unitary operation applied 
to a single qubit.

In what follows, we derive a Lindblad master equation~\cite{Lin76,GoKoSu76}, 
which generates the quantum map $\Lambda(\bm A,\vec b\, )$ when integrated over 
a fixed time interval. To do so, we use the divisibility of $f$, which implies 
that for any finite $n > 0$ there exist a distribution $h_n$ such that the 
ensemble $\{ U \}_f$ is indistinguishable from 
$\{ u_n\, \ldots u_2\, u_1 \}_{h_n}$. Then, we introduce $t= j/n$, with 
$0\le j \le n$ and $\delta = 1/n$, to find
\begin{align}
&\varrho_t= \langle u_j\, \ldots u_1\, \varrho_0\, u_1^\dagger\, \ldots 
u_j^\dagger \rangle\; , \quad \varrho_1 = \Lambda[\varrho_0]\; ,
\notag\\ 
&\text{and}\quad \varrho_{t+\delta}= 
\langle u_{j+1}\, \varrho_t\, u_{j+1}^\dagger\rangle\; ,
\end{align}
due to the statistical independence of the $\{ u_j\}$. Here, again, we use the 
notation $\la\, \cdots\,\ra$ for the average over the distribution $h_n$.
Thereby, we arrive at the differential equation \cite{GoKoSu76},
\begin{equation}
\frac{d}{dt}\varrho = \lim_{n\to\infty} \frac{1}{\delta}\big ( \,
\langle u\, \varrho\, u^\dagger\rangle -\varrho \big ) \; .
\end{equation}
We use once more the fact that $h_n$ is ever more localized at the identity, 
and insert for $u= D^{(s)}(g)$ the exponential map in Eq.~(\ref{aN:expomap}), 
expanding it up to second order, which yields 
\begin{align}
&\langle u\, \varrho\, u^\dagger\rangle_{h_n} \approx \big\langle \big[ \One - 
   i\, \vec n\cdot\vec{L}^{(s)} - {\textstyle \frac{1}{2}}\, 
   (\vec n\cdot\vec{L}^{(s)})^2\big ]\, \varrho\, \notag\\
&\qquad\qquad\qquad\qquad \times 
   \big [ \One + i\, \vec n\cdot\vec{L}^{(s)} - 
   {\textstyle \frac{1}{2}}\, 
   (\vec n\cdot\vec{L}^{(s)})^2\big ]\, \big\rangle_{h_n} \notag\\
&\quad \approx \varrho - \big\langle 
   i\, [\vec n\cdot\vec{L}^{(s)} , \varrho] 
   + {\textstyle \frac{1}{2}}\, \{ (\vec n\cdot\vec{L}^{(s)})^2 , \varrho \}  
\notag\\
&\quad\qquad -\, \vec n\cdot\vec{L}^{(s)}\, \varrho\, 
   \vec n\cdot\vec{L}^{(s)} \big\rangle_{h_n}  \notag \\
&\quad = \varrho - i\,  [\vec b \cdot\vec{L}^{(s)} , \varrho]\notag \\
&\quad\qquad -\, \frac{1}{2}\sum_{i,j} A_{ij}\, 
   \big ( \, \{ L_i^{(s)} L_j^{(s)} , \varrho \} 
   -2\,  L_i^{(s)}\varrho\, L_j^{(s)} \, \big )\; ,
\end{align}
where we have used commutators $[\, \cdot , \cdot\, ]$ and anti-commutators
$\{ \, \cdot , \cdot\, \}$ to shorten the notation. Thereby we finally arrive 
at the desired master-equation,
\begin{align}
\frac{d}{dt}\varrho &= -i\, [\vec b\cdot\vec{L}^{(s)} , \varrho]
   + \mathcal{D}[\varrho]\; , \notag\\
\mathcal{D}[\varrho] &= \sum_{i,j} A_{ij}\, 
   \Big (\, L_i^{(s)}\varrho\, L_j^{(s)}
   - \frac{1}{2} \{ L_i^{(s)} L_j^{(s)} , \varrho \} \, \Big )\; .
\label{ME:MasterEqGen}\end{align}
In the finite-dimensional case, and for unital quantum processes
this is the most general form of a Lindblad master equation, where it is 
required that $\bm A$ is real and positive semi-definite, and where $\vec b$ 
determines the Hamiltonian of the system~\cite{GoKoSu76,BrePet02}. For $t=1$, 
we obtain the quantum channel defined in Eq.~(\ref{generalnormalmap}). For 
other values $t \ge 0$, we obtain a one-parameter family $\Lambda_t$ of quantum 
channels, i.e. a quantum process. In view of the construction, it is clear 
that this process is CP-divisible (i.e. divisible in completely positive and
trace preserving maps) in the sense of Ref.~\cite{MDG19}. 

In Ref.~\cite{DaZiPi19}, the authors define a quantum channel to be 
``Lindblad divisible'' (L-divisible), if there exits a Lindblad master equation 
(with time-independent coefficients) which generates that channel in a finite 
amount of time. In this sense, normal quantum channels are always L-divisible, 
and at the same time, any L-divisible unital quantum channel is normal.
This means that any Lindblad master equation can be solved (numerically) by an
appropriate diffusive random walk in the corresponding dynamical group. This is
precisely the basis of the so called ``quantum state diffusion'' 
method~\cite{NGisin_1992,Castin1993,percival1998quantum}.

It is however possible that a given normal quantum channel is generated by 
different normal distributions and hence different Lindblad master equations, 
both objects parameterized by different tuples $(\bm A, \vec b)$. We will
analyze this situation for the single qubit case in the following Sec.~\ref{S}.

\section{\label{S} Single qubit normal quantum channels}

Here, we investigate the application of normal distributions for describing 
imperfect quantum gates acting on single qubits (two-level quantum systems). 
This means that instead of a fixed unitary transformation $U_0\in {\rm SU}(2)$,
we assume that a random transformation with a normal distribution in 
${\rm SU}(2)$ is applied; see Eq.~(\ref{generalnormalmap}). The result is a 
non-unitary quantum operation, {\it i.e.} a linear, completely positive and 
trace-preserving (CPTP)~\cite{NieChu00} map, or simply a quantum channel, in 
the space of $2$$\times$$2$ density matrices.

The special unitary group $SU(2)$ has the Wigner D-matrices $D^{(s)}(g)$ as 
irreducible representations with $s = 0,1/2,1,3/2,\ldots$. It has $r=3$ 
generators which are the angular momentum operators divided by $\hbar$. 
Since we are dealing with qubits, $U(g)= D^{(1/2)}(g)$, where the generators 
can be written in terms of the Pauli matrices
$\vec L^{(1/2)} = \vec\sigma/2$~\cite{Sakurai94}. The normal distribution 
$f(g)$ is parameterized by the pair $(\bm A,\vec b\, )$ [see 
Eq.~(\ref{aN:parametrization})].

\subsection{\label{SC} Choi-matrix representation}

In what follows, we will calculate the Choi-matrix 
representation~\cite{BenZyc06,HeiZimBook11} for the CPTP map 
$\Lambda(\bm A,\vec b\, )$ given as in Eq.~(\ref{generalnormalmap}).
\begin{align}
\mathcal{C}(\bm A,\vec b\, ) &= \sum_{i,j=0}^1 |i\rangle\langle j| \otimes 
   \Lambda[\, |i\rangle\langle j|\, ] \notag\\
 &= \int_{{\rm SU}(2)}dg\; f(g)\; \mathcal{C}[D^{(1/2)}(g)]\; ,
\label{SC:defChoiDist}\end{align}
with
\begin{align}
&\mathcal{C}[D^{(1/2)}(g)] = \sum_{i,j=0}^1 |i\rangle\langle j| \otimes
   D^{(1/2)}(g)\, |i\rangle\langle j|\, D^{(1/2)}(g)^\dagger\; ,
\label{SC:defChoiElements}\\
&D^{(1/2)}(g) = R_z(\alpha)\, R_y(\beta)\, R_z(\gamma) \notag\\
&\quad = \begin{pmatrix} e^{-i\, (\alpha + \gamma)/2}\, \cos(\frac{\beta}{2}) &
   -\, e^{-i\, (\alpha - \gamma)/2}\, \sin(\frac{\beta}{2})\\[3pt]
   e^{\, i\, (\alpha - \gamma)/2}\, \sin(\frac{\beta}{2}) & 
   e^{\, i\, (\alpha + \gamma)/2}\, \cos(\frac{\beta}{2})\end{pmatrix}\; ,
\end{align}
using Euler angles and corresponding rotation matrices~\cite{Sakurai94}.

The product of matrix elements of the Wigner D-matrices in 
Eq.~(\ref{SC:defChoiElements}) can be written as linear combinations of the 
matrix elements of single Wigner D-matrices with $s=0$ and $s=1$. This is done 
in the Appendix, with the result given in Eq.~(\ref{AppA:Res}). 
For $s=0$, the only matrix element is $D^{(0)}(g) = 1$, and according to 
Eq.~(\ref{aNH:DefFouInt}), $\mathcal{F}_0[f] = 1$. For $s=1$, 
\begin{equation}
\mathcal{F}_1[f] = \int dg\; f(g)\; D^{(1)}(g) = e^{-\mathcal{M}_1}\; ,
\end{equation}
where
\begin{align}
\mathcal{M}_1 &= \frac{1}{2}\begin{pmatrix} 
   A_{33} + \frac{A_{11} + A_{22}}{2} & \frac{A_{13} - i A_{23}}{\sqrt{2}} &
   -i A_{12} + \frac{A_{11} - A_{22}}{2}\\
   \frac{A_{13} + i A_{23}}{\sqrt{2}} & A_{11} + A_{22} &
   \frac{-A_{13} + i A_{23}}{\sqrt{2}}\\
   i A_{12} + \frac{A_{11} - A_{22}}{2} & \frac{-A_{13} -i A_{23}}{\sqrt{2}} &
   A_{33} + \frac{A_{11} + A_{22}}{2}\end{pmatrix} \notag\\
&\qquad - \begin{pmatrix} 
   -i b_3 & -\, \frac{b_2 + i b_1}{\sqrt{2}} & 0\\
   \frac{b_2 -i b_1}{\sqrt{2}} & 0 & -\, \frac{b_2 + i b_1}{\sqrt{2}}\\
   0 & \frac{b_2 -i b_1}{\sqrt{2}} & i b_3\end{pmatrix}\; .
\label{IrRepGen}\end{align}
As a result we find that the Choi-matrix elements in Eq.~(\ref{SC:defChoiDist}) 
are given in terms of the matrix $e^{-\mathcal{M}_1}$. Hence, in general, this
requires to compute the eigenvalues and corresponding eigenvectors of a 
non-Hermitian $3$$\times$$3$ matrix. This means that in principle, it is 
possible though very cumbersome to obtain analytical results. In this paper, we 
therefore calculate these matrix elements either numerically or consider 
simpler special cases.

\subsection{\label{ME} Lindblad master equation}

In terms of the Pauli matrices,
$\vec\sigma = 2\, \vec L^{(1/2)}$, the master equation for a single qubit 
normal channels is 
\begin{align}
\frac{d}{dt}\varrho &= -\, \frac{i}{2}\, [\vec b\cdot\vec\sigma , \varrho]
   + \mathcal{D}[\varrho]\; , \notag\\
\mathcal{D}[\varrho] &= \frac{1}{4}\sum_{i,j=1}^3 A_{ij} \Big (
   \sigma_i\, \varrho\, \sigma_j - \frac{1}{2}\, 
   \{ \sigma_i \sigma_j , \varrho\}\, \Big )\; .
\label{ME:MasterEq}\end{align}
Since $\bm A$ is real, every normal quantum channel defines a 
unital Lindblad-divisible (L-divisible) quantum channel in the sense of 
Ref.~\cite{DaZiPi19} and every unital L-divisible channel can be associated 
with one or several normal quantum process(es).

\subsection{\label{SU} Pauli representation}

In the previous Sec.~\ref{SC}, we discussed the Choi-matrix representation of
a normal quantum channel. Here, it will prove convenient to introduce the 
representation in the Pauli basis. Since every normal map 
$\Lambda(\bm A, \vec b\, )$ (we may eventually skip its arguments) is unital, 
we have
\begin{equation}
\varrho = \frac{\One + \vec n\cdot\vec\sigma}{2} \quad\to\quad 
\varrho' = \Lambda[\varrho] = \frac{\One + \vec n'\cdot\vec\sigma}{2}\; ,
\end{equation}
such that 
\begin{equation}
\vec n' = \mathcal{R}\, \vec n\; , \qquad \mathcal{R}_{ij} = \frac{1}{2}\,
   {\rm tr}\big ( \sigma_i\, \Lambda[\sigma_j]\, \big )\; .
\label{SU:PauliRep}\end{equation}
where $\vec n$ and $\vec n'$ are vectors in the Bloch sphere and
$\mathcal{R}$ is a real, in general non-symmetric matrix with non-negative 
singular values less or equal to one \cite{BenZyc06}. 

For the representation $\mathcal{R}$ the composition of two maps corresponds to 
the matrix product of their representations. Therefore, we obtain from the 
master equation~(\ref{ME:MasterEq}) with $\varrho = \Lambda_t[\varrho_0]$
\begin{align}
\frac{d}{dt}\, \mathcal{R}(t) &= \mathcal{L}\, \mathcal{R}(t)
\quad\Rightarrow\quad
\mathcal{R} = \mathcal{R}(1) = e^\mathcal{L}\; , \notag\\
\mathcal{L}_{ij} &= \frac{1}{2}\, {\rm tr}\Big [ \sigma_i\, \Big ( -\frac{i}{2}
   [\vec b \cdot \vec\sigma , \sigma_j] + \mathcal{D}[\sigma_j] \Big ) 
   \, \Big ]\; .
\end{align}
Specifically, we obtain
\begin{equation}
\mathcal{L} = \frac{1}{2}\big ( \bm A - {\rm tr}(\bm A)\, \One\big ) 
+ [ \vec b \,]_\times\; .
\label{PauliRepGen}\end{equation}
where $[ \vec b \,]_\times$ is the cross product matrix of vector $\vec{b}$ 
defined as:
\begin{equation}
[ \vec b \,]_\times =\begin{pmatrix}
	0 & -b_3 & b_2\\
	b_3 & 0 & -b_1\\
	-b_2 & b_1 & 0 \end{pmatrix}.
\end{equation}
With Eq.~(\ref{IrRepGen}), which gives the generator of a normal quantum 
channel in terms of the Peter-Weyl basis, and Eq.~(\ref{PauliRepGen}), we have 
two alternatives for calculating the quantum channel corresponding to a certain
normal distribution. But note that both options are unitarily equivalent.
Indeed,
\begin{equation}
U\; \big ( - \mathcal{M}_1\big )\; U^\dagger = \mathcal{L}\; ,  \qquad 
U= \frac{1}{\sqrt{2}}\begin{pmatrix}
	-i&0&i\\
	1&0&1\\
	0& \sqrt{2}i &0 \end{pmatrix}\; .
\end{equation}

\subsection{Non-uniqueness}

We already know that for any given unital, L-divisible quantum channel, there
exists a normal distribution with diffusion matrix $\bm A$ and drift-vector 
$\vec b$, which produces this channel. Here, we ask whether this distribution 
is unique, and if not how many and what are the different normal distributions 
which generate the same channel. This is a striking difference to normal 
distributions in a flat space, where the knowledge of the first and the second 
moments of the distribution uniquely define the normal distribution in 
question. 

In order to study the (lack of) uniqueness, we consider the space of normal 
distributions parameterized according to Eq.~(\ref{aN:parametrization}) by the 
tuple $(\bm A, \vec b\, )$. In this space, we may consider the equivalence 
relation $(\bm A, \vec b\, ) \sim(\bm A', \vec b')$ meaning that the two 
distributions $(\bm A, \vec b)$ and $(\bm A', \vec b')$ produce the same 
quantum channel. Then we denote by $[(\bm A, \vec b)]$ the equivalence class of 
all those distributions which produce the same quantum channel.

The following considerations are closely related to the investigation of 
L-divisibility in Refs.~\cite{DaZiPi19} and we use some results on 
matrix-logarithms from \cite{CulverLogaritmo}. However, before doing so, the 
following observation will prove useful: Let us denote with
$\mathcal{R}(\bm A, \vec b)$ the quantum channel produced by the normal 
distribution $(\bm A, \vec b)$, as we are using the Pauli matrix representation
introduced in Eq.~(\ref{SU:PauliRep}). Then it holds, for any 
$O\in {\rm SO}(3)$:
\begin{equation}
\bm O\, \mathcal{R}(\bm A, \vec b)\, \bm O^T = 
\mathcal{R}(\bm O\bm A\bm O^T, \bm O\, \vec b)\; .
\label{OrthogNormalR}\end{equation}
This suggests to consider orthogonal transformations in the space of normal
distributions, plainly denoted as 
$\bm O\, (\bm A, \vec b)\, \bm O^T = (\bm O\bm A\bm O^T, \bm O\, \vec b)$.
Then Eq.~(\ref{OrthogNormalR}) implies that 
\begin{equation}
(\bm A, \vec b) \sim (\bm A', \vec b') \;\Leftrightarrow\;
\bm O (\bm A, \vec b)\, \bm O^T \sim 
\bm O (\bm A', \vec b')\, \bm O^T\; ,
\end{equation}
and therefore
\begin{equation}
  [ (\bm O\bm A \bm O^T, \bm O\, \vec b) ] =
\bm O\, [ (\bm A, \vec b)]\, \bm O^T \; .
\end{equation}
Hence, since we can always find an orthogonal transformation, which 
diagonalizes $\bm A = \bm O\, \bm\lambda\, \bm O^T$, the following study of 
equivalence classes can be limited to normal channels, where $\bm A$ is 
diagonal. 

\begin{figure}
\includegraphics[width= 0.5\textwidth]{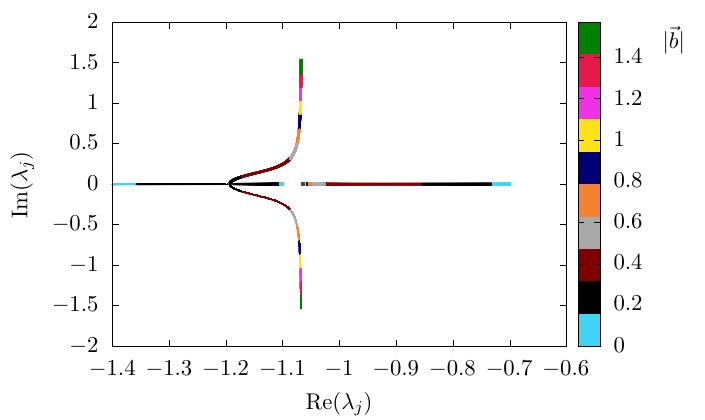}
\caption{Level dynamics of the eigenvalues $\lambda_j$ of $\mathcal{L}$ in the 
complex plane as the magnitude $|\vec b|$ of the drift vector is increased for 
a fixed, diagonal diffusion matrix ${\bm A}$. }
\label{f:evaltrace}\end{figure}

In figure \ref{f:evaltrace} we show the typical behavior of the eigenvalues of 
$\mathcal{L}$, when the magnitude $|\vec b|$ of the drift vector is 
increased from zero to $1.5$. We assume $\bm A$ to be diagonal and choose
a random orientation for $\vec b$. We use a discrete color-scale to visualize
the movement of the eigenvalues in the complex plane as a function of 
$|\vec b|$. As long as $|\vec b|$ is small enough, the eigenvalues of 
$\mathcal{L}$ remain real. For $|\vec b| \approx 0.2$ we reach a point, where
two eigenvalues (say $\lambda_1$ and $\lambda_2$) become equal and then head 
off into the complex plane. Beyond that point, $\lambda_3$ remains real, while
$\lambda_2 = \lambda_1^*$. This is the typical behavior of the eigenvalues of
a $\mathcal{P}\mathcal{T}$-symmetric 
Hamiltonian ~\cite{valtierra2021pt,BenderPTSymmetry2019}, where 
$\lambda_1 = \lambda_2$ at a so-called ``exceptional point''. At this point the 
corresponding eigenvectors become linearly dependent and hence $\mathcal{L}$ 
defective. For large $|\vec b|$, finally, all eigenvalues align on a line,
parallel to the imaginary axis. 

In what follows we discuss two cases: (i) all eigenvalues are real, and (ii) 
one eigenvalue is real and the other two are complex-conjugated. Subsequently,
case (i) is further divided into the cases with and without drift, and case 
(ii) is further divided into the cases where the drift matrix 
$[\vec b\, ]_\times$ commutes or not with the diffusion matrix $\bm A$. 

\paragraph{Only real eigenvalues; no drift.}
This means that we are treating a diagonal Pauli channel, which is most 
intuitively defined in terms of the Kraus operator 
representation~\cite{KRAUS1971311,BenZyc06,HeiZimBook11}:
\begin{align}
\Lambda_{\rm P}[\varrho] &= (1 -p_1 -p_2 -p_3)\, \varrho \notag\\
 &\quad + p_1\, \sigma_x \varrho \sigma_x + p_2\, \sigma_y \varrho \sigma_y
   + p_3\, \sigma_z \varrho \sigma_z\; ,
   \label{PauliChannel}
\end{align}
where $p_j \ge 0$ and $\sum_j p_j \le 1$. This is a unital channel with the 
Pauli representation
\begin{equation}
\mathcal{R}_{\rm P} = \One - 2\begin{pmatrix} p_2+p_3 & 0 & 0\\
   0 & p_1+p_3 & 0\\ 0 & 0 & p_1+p_2\end{pmatrix}\; .
\label{PauliPRep}\end{equation}
For a normal channel with diagonal matrix $\bm A$ and $\vec b = 0$, we 
find that $\mathcal{R} = e^{\mathcal L}$ becomes diagonal with its elements 
equal to $\mathcal{R}_{jj} = e^{- ({\rm tr}(\bm A) - A_{jj})/2}$. Comparing 
this result with Eq.~(\ref{PauliPRep}), leads to a linear system of equations 
for the parameters $\{ p_j\}$ with the solution
\begin{equation} 
p_j = \frac{1 - e^{-{\rm tr}(A)/2}\sum_{k=1}^3 (-1)^{\delta_{jk}} 
   e^{A_{kk}/2}}{4} \; .
\label{parametrosPauli}\end{equation}
According to Ref.~\cite{CulverLogaritmo}, the equivalence class 
$[(\bm A,\vec b)]$ has only one element in this case. Note that 
$0\le p_j \le 1/4$, where $p_j = 1/4$ corresponds to the limit 
$A_{jj} \to \infty$, where the distribution $f(g)$ converges to the uniform
distribution in SU$(2)$. In this case we obtain the completely depolarizing
channel, which maps any quantum state to the uniform mixture 
$\varrho = \One/2$.

\paragraph{Real eigenvalues with drift}
If we have a drift term and real eigenvalues, the drift term cannot commute 
with $\bm A$. This case is illustrated in Fig.~\ref{f:evaltrace} when the 
magnitude of the drift vector $\vec b$ is still sufficiently small. Since the 
generator $\mathcal{L}$ and therefore $\mathcal{R}$ have only real eigenvalues, 
the equivalence class  $[(\bm A,\vec b)]$ has only one 
element~\cite{CulverLogaritmo}.

\paragraph{Complex eigenvalues, commuting drift.}
The drift matrix $[\vec b\, ]_\times$ can only commute with $\bm A$ if at least 
two elements of $\bf A$ are equal. In this case, there exists a common 
ortho-normal basis in which, both, $\bm A$ and $[\vec b\, ]_\times$ are 
diagonal. Since $[\vec b\, ]_\times$ is anti-symmetric it has only imaginary
eigenvalues, so that due to Eq.~(\ref{PauliRepGen}) the eigenvalues of 
$\mathcal{L}$ must be complex.

Assume for instance that $A_{11} = A_{22}$, 
then the drift term commutes with $\bf A$ iff $\vec b = (0,0,b_3)$. As a 
consequence the normal channels which correspond to the drift alone and the 
diffusion alone, also commute, i.e 
$\Lambda(\bm A, \vec b) = \Lambda(\bm o, \vec b) \circ \Lambda(\bm A, \vec o)$ 
becomes the combination of a unitary rotation and a Pauli channel, which can be 
applied in any order. Here, the equivalence class $[({\bm A}, \vec b)]$ has an 
infinite number of elements, which are all distributions with parameters 
${\bm A} = {\rm diag}(A_{11},A_{11},A_{33})$ and $b_3 = b'_3 + 2 \pi k$, with
$k \in \mathbb{Z}$.   

\begin{figure}
\includegraphics[width=0.5\textwidth]{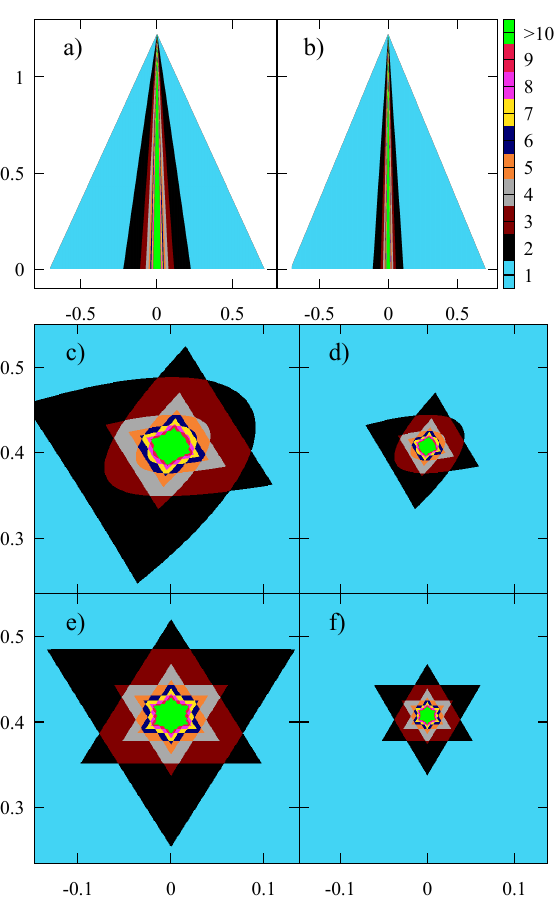}
\caption{Number of normal distributions (indicated by different colors)
generating the same quantum channel as a function of the diagonal $\bm A$ shown 
in the plane ${\rm Tr}\left({\bm A}\right)=1$. For the drift term, we choose 
different directions $\vec b \propto {\rm e}_z$ [panels a), b)], 
$\propto ({\rm e}_y + {\rm e}_z)$ [panels c), d)], and 
$\propto ({\rm e}_x + {\rm e}_y + {\rm e}_z)$ [panels e), f)]. In the panels on
the left, $\|\vec b\| = 1$ in the panels on the right, $\|\vec b\| = 1/2$. }
\label{triangulo}\end{figure}

\paragraph{Complex eigenvalues; non-commuting drift}
This case is also illustrated in Fig.~\ref{f:evaltrace}, for $\|\vec b\|$ 
sufficiently large. This is the most interesting case, where several normal
distributions may exist, which result in the same quantum channel. To search
for such distributions, we calculate systematically real matrix-logarithms
of $\mathcal{R}$. Since the logarithm must be real, two eigenvalues of 
$\mathcal{R}$ must be complex conjugates and the other one must be 
real~\cite{CulverLogaritmo}. Diagonalizing $\mathcal{R}$ as 
$\mathcal{R} = P D P^{-1}$, where $D={\rm diag}\left(d_1,d_1^*,d_3\right)$ and 
$d_3 \in \mathbb{R}$ the real matrix-logarithms can be written as 
\begin{equation}
\ln\left(\mathcal{R} \right) = P \begin{pmatrix}
  \ln(d_1) + 2\pi ik &0&0\\ 0& \ln(d_1^*) - 2\pi ik & 0\\
  0 & 0 &\ln(d_3)\\ \end{pmatrix} P^{-1}\, ,
\end{equation}
where $k \in \mathbb{Z}$. Then we consider 
$\ln\left(\mathcal{R} \right)=\mathcal{L}$ and solve for $\bm A$ and $\vec b$ 
as defined in  Eq.~(\ref{PauliRepGen}), accepting only those solutions, where 
$\bm A$ has non-negative eigenvalues.

In the case, the distributions which yield the same quantum channel differ in 
the drift term and in the diffusion matrix in a complicated way. For 
illustration, we show in Fig.~\ref{triangulo} the number of different normal 
distributions contained in an equivalence class $[(\bm A,\vec b)]$ as a 
function of the diagonal $\bm A$ and for different drift vectors $\vec b$. For 
that purpose, we use a false-color plot in the plane 
$A_{11} + A_{22} + A_{33} = 1$, such that the region of allowed values for 
$\bm A$ is given by an equilateral triangle with sides of length $\sqrt{2}$.

In the panels a) and b), the drift vector is parallel to the $z$-axis, which is 
a particularly simple case, such that the number of equivalent distributions
$\mathcal{N}_{\rm equil}$ can be calculated exactly. In this case,
$\mathcal{N}_{\rm equil}$ increases inasmuch as one approaches the vertical 
line $A_{11} = A_{22}$, while on that line $\mathcal{N}_{\rm equil} = \infty$. 
In the panels c), d), e), and f) the number of equivalent distributions is
calculated numerically up to a maximum number of twelve.

When reducing the magnitude of the drift vector, the multiplicity of equivalent
distributions is reduced. This is in line with the observation made in 
Fig.~\ref{f:evaltrace} that by reducing the magnitude of $\vec b$, one 
eventually arrives in a regime where the eigenvalues of $\mathcal{L}$ are all 
real, so that the multiplicities disappear.

\section{\label{T} Two qubit normal channels and correlated errors}

In this section we construct normal quantum channels, which can describe
correlated noise affecting two-qubit quantum gates in a general, intuitive 
manner. For that purpose, we consider normal distributions in the group 
${\rm SU}(2)\otimes {\rm SU}(2)$ and compare the resulting quantum channels to 
a model of correlated Pauli 
errors~\cite{CorrelatedCaruso,CorrelatedMachiavelo}.

\subsection{\label{TG} Choi-matrix representation}

Following the methodology outlined in Sec.~\ref{aN}, we note that 
${\rm SU}(2)\otimes {\rm SU}(2)$ is a compact Lie group with the following
six generators:
\begin{equation}
\vec L^{(s_1,s_2)} = \begin{pmatrix}
L_x^{(s_1)} \otimes \One \\
L_y^{(s_1)} \otimes \One \\
L_z^{(s_1)} \otimes \One \\
\One \otimes L_x^{(s_2)} \\
\One \otimes L_y^{(s_2)} \\
\One \otimes L_z^{(s_2)} \end{pmatrix}\; .
\end{equation}
The irreducible representations of this group are given by tensor products of 
the Wigner D-matrices,
\begin{equation}
D^{(s_1,s_2)}(g) = D^{(s_1)}(g_1)\otimes D^{(s_2)}(g_2)\; , 
\end{equation}
with $g= (g_1,g_2)$ and $g_1,g_2 \in {\rm SU}(2)$. Therefore, the family of 
normal distributions is parameterized by the diffusion matrix $\bm A$ of 
dimension $6$$\times$$6$, and the drift vector $\vec b$ of the same dimension. 
Then Eq.~(\ref{aN:parametrization}) becomes
\begin{align}
&\int_{{\rm SU}(2)} dg_1 \int_{{\rm SU}(2)} dg_2\; f(g)\;  D^{(s_1,s_2)}(g) 
 = \exp\big ( - \mathcal{M}_{s_1,s_2}\big )\notag\\
&\mathcal{M}_{s_1,s_2} =
  \frac{1}{2}\, \vec L^{(s_1,s_2)}{}^T \cdot \bm A \cdot
      \vec L^{(s_1,s_2)} + i\, \vec{b}\cdot\vec{L}^{(s_1,s_2)}\; .
\label{su2xsu2distribution}\end{align}
In order to calculate the Choi-matrix representation of the induced quantum 
channel, we note that
\begin{align}
&\mathcal{C}(\bm A,\vec b\, ) = \sum_{i,j,k,l=0}^1 |ij\rangle\langle kl| 
   \otimes \Lambda[\, |ij\rangle\langle kl|\, ] \notag\\
&\quad = \int_{{\rm SU}(2)\times {\rm SU}(2)}dg_1\, dg_2\; f(g)\; 
   \mathcal{C}[D^{(1/2)}(g_1)\otimes D^{(1/2)}(g_2)]\; ,
\label{TG:defChoiDist}\end{align}
where
\begin{align}
&\mathcal{C}[D^{(1/2)}(g_1)\otimes D^{(1/2)}(g_2)]\notag\\
&\qquad = \sum_{i,j,k,l=0}^1 
   |ij\rangle\langle kl| \otimes
   D^{(1/2)}(g_1)\, |i\rangle\langle k|\, D^{(1/2)}(g_1)^\dagger\notag\\
&\qquad\qquad\qquad\qquad\qquad\quad\otimes
   D^{(1/2)}(g_2)\, |j\rangle\langle l|\, D^{(1/2)}(g_2)^\dagger\notag\\
&\qquad \simeq 
\mathcal{C}[D^{(1/2)}(g_1)] \otimes \mathcal{C}[D^{(1/2)}(g_2)]\; .
\label{TG:defChoiElements}\end{align}
 We write ``$\simeq$'' because to arrive at the last line, one has to ``swap'' 
the second with the third factor in the fourfold tensor product. This shows 
that for the Choi-matrix representation we will need the Fourier matrix 
coefficients from (i) $\mathcal{M}_{0,1}$, (ii) $\mathcal{M}_{1,0}$, and
(iii) $\mathcal{M}_{1,1}$. The first two cases correspond to the Fourier matrix
coefficients of:
\begin{equation} 
\text{(i)}\; \int_{{\rm SU}(2)} dg_1\; f(g_1,g_2)\; , 
   \quad\text{and}\quad
 \text{(ii)}\; \int_{{\rm SU}(2)} dg_2\; f(g_1,g_2)\; , 
\end{equation}
which are normal distributions in ${\rm SU}(2)$ and have been treated in 
Sec.~\ref{SC}. For instance, in the case (i)
\begin{equation} 
\vec L^{(0,1)}{}^T\cdot \bm A \cdot \vec L^{(0,1)}
 = \vec L^{(1)}{}^T\cdot \begin{pmatrix} A_{44} & A_{45} & A_{46}\\
      A_{45} & A_{55} & A_{56}\\
      A_{46} & A_{56} & A_{66}\end{pmatrix} \cdot \vec L^{(1)} \; . 
\end{equation}
Case (iii) represents the new degree of difficulty, where the 
matrix $\mathcal{M}_{1,1}$ is of dimension $9$$\times$$9$.

\subsection{\label{ME2} Lindblad master equation}

The resulting master equation is of the same form as in 
Eq.~(\ref{ME:MasterEqGen}) and can be written as
\begin{align}
\frac{d}{dt}\varrho &= -i\, [\vec b\cdot\vec{L}^{(s_1,s_2)} , \varrho]
  + \mathcal{D}[\varrho]\; , \quad s_1=s_2=\frac{1}{2} 
\label{ME2:MasterEq2}\\
\mathcal{D}[\varrho] &= \sum_{i,j=1}^6 A_{ij} 
  \Big [ L_i^{(s_1,s_2)}\varrho L_j^{(s_1,s_2)} - \frac{1}{2} 
  \big\{ L_i^{(s_1,s_2)} L_j^{(s_1,s_2)} , \varrho \big\} \Big ]\; .
\notag\end{align}
Following the derivation in Sec.~\ref{SU}, we write the generator 
$\mathcal{L}^{(2)}$ for the normal channels associated to the normal 
distributions defined in Eq.~(\ref{su2xsu2distribution}) in the Pauli basis 
\begin{equation}
\frac{1}{2}\{ \sigma_j\otimes\sigma_k \}_{0\le j,k\le 3}\; , \qquad 
   \sigma_0 = \One\; ,
\label{base2}\end{equation}
in terms of the parameters $\bm A$ and $\vec{b}$ of the distribution. To this
end, we first define 
\begin{equation}
{\bm A}= \begin{pmatrix}
	{\bm A}_1 & {\bm F}\\
	{\bm F}^t & {\bm A}_2
\end{pmatrix}, \qquad \vec{b} = \begin{pmatrix}
\vec{b}_1\\
\vec{b}_2\\
\end{pmatrix},
\label{ME2:NormParam}\end{equation}
where ${\bm A}_1$, ${\bm A}_2$ and ${\bm F}$ are $3$$\times$$3$ matrices and 
the vectors $\vec{b}_{1,2}$ have three entries each. 
With this notation, the generator can be written as
\begin{equation}
\mathcal{L}^{(2)}= 0  \oplus \mathcal{L}({\bm A}_1,\vec{b}_1)\oplus 
   \mathcal{L}({\bm A}_2,\vec{b}_2) \oplus {\bm M}\; ,
   \label{GeneradorLambda2}
\end{equation}
where $\mathcal{L}({\bm A},\vec{b})$ is the single-qubit normal quantum channel 
generator defined by the parameters $\bm A$ and $\vec{b}$ and the 
$9$$\times$$9$ matrix ${\bm M}$ is defined as
\begin{equation}
{\bm M} = \begin{pmatrix}
{\bm m}_{11} & {\bm m}_{12}& {\bm m}_{13}\\  
{\bm m}_{21} & {\bm m}_{22}& {\bm m}_{23}\\
{\bm m}_{31} & {\bm m}_{32}& {\bm m}_{33}\\  
\end{pmatrix}\; ,
\end{equation}
with
\[  {\bm m}_{ij} = \begin{cases}
    \left(\mathcal{L}({\bm A}_1,\vec{b}_1)_{ii} \right) \One 
      + \mathcal{L}({\bm A}_2,\vec{b}_2)  & : i=j\\
    \left(\mathcal{L}({\bm A}_1,\vec{b}_1)_{ij} \right) \One 
      + (-1)^{i+j+1}\left[\vec{F}_{6-i-j}\right]_x & : i>j\\
    \left(\mathcal{L}({\bm A}_1,\vec{b}_1)_{ij} \right) \One 
      + (-1)^{i+j}\left[\vec{F}_{6-i-j}\right]_x & : i<j
\end{cases}\; , \]
where $\vec{F}_n$ denotes the $n$'th column vector in ${\bm F}$.

\subsection{\label{TI} Correlated Pauli errors}

In the literature qubit errors and correlated qubit errors are usually modeled
by Pauli channels. In the most general case, a correlated Pauli error for two
qubits may be defined as~\cite{CorrelatedCaruso,CorrelatedMachiavelo}
\begin{equation}
\Lambda_{\rm cP}[\varrho]= \sum_{i,j = 0}^3 p_{ij}\; 
   (\sigma_i\otimes \sigma_j)\, \varrho\, (\sigma_i\otimes \sigma_j)\; ,
\label{correlatedPauli}\end{equation}
where $\sum_{i,j} p_{ij} = 1$. According to this model, the two qubits undergo
unitary Pauli-rotations $\sigma_i\otimes \sigma_j$ with probability $p_{ij}$.

A special case of the quantum channel $\Lambda_{\rm cP}$ describes the case of
a weighted mixture of applying either the same Pauli error (maximally 
correlated) or independent Pauli errors to the two qubits. In this case, 
\begin{equation}
p_{ij}= (1-m)\; p_i\, p_j + m\; p_i\, \delta_{ij}\; , 
\end{equation}
with $p_i\geq 0$, $\sum_{i}p_i=1$, and $0 \le m \le 1$ denoting the probability 
to apply the same Pauli error to both qubits. 

For a general $0 < m < 1$ this quantum channel is a convex mixture of two Pauli 
channels -- one where the Pauli errors are applied independently to both qubits 
($m=0$), and one where the same Pauli error is applied to both qubits ($m=1$). 
In Ref.~\cite{ConvexMixMarkov} it has been shown that such a 
mixture may be non-Markovian. In this case it could not be the result of an evolution according to a Lindblad master equation. 

For another special case, we first generalize the model above to allow for 
different error probabilities in both qubits:
\begin{equation}
p_{ij} = (1-m)\; p_i\, q_j + m\; \frac{p_i + q_i}{2}\; \delta_{ij}\; ,
\end{equation}
with $\sum_i p_i = \sum_j q_j = 1$. Then we make the errors isotropic, such 
that
\begin{align}
p_1 &= p_2 = p_3 = p\; , \qquad p_0 = 1-3p\; , \notag\\
q_1 &= q_2 = q_3 = q\; , \qquad q_0 = 1-3q\; .
\end{align}
Again, the resulting quantum channel is Lindblad divisible and hence Markovian 
for $m=0$ and $m=1$, but probably not for values in between. We will denote 
this quantum channel by $\Lambda_{\rm cP}(m)$. In the Pauli basis, cf. 
Eq.~(\ref{PauliPRep}), this channel can be written as
\begin{equation}
\mathcal{R}^{(2)}_{\rm cP} = 1\oplus \left(\mathcal{R}_{\rm P}(p) 
   + \bm c \right) \oplus \left( \mathcal{R}_{\rm P}(q) - \bm c \right) 
   \oplus \bm W_{\rm cP}\; ,
\end{equation}
where $\mathcal{R}_{\rm P}(p) = (1-4p)\, \One$, $\bm c = 2m(p-q)\One$, and
\begin{equation}
\bm W_{\rm cP} =w_1 \oplus w_2\One_{\rm 3 \times 3} 
   \oplus w_1 \oplus w_2\One_{\rm 3 \times 3} \oplus w_1 \; ,
\label{TI:WcP}\end{equation}
with $w_1 = 1 + 4(m-1)(p+q(1-4p))$ and $w_2 = w_1 -2m(p+q)$.

\subsection{\label{TC} Correlated normal errors}

We believe that two-qubit quantum channels derived from normal distributions of
the form given in Eq.~(\ref{su2xsu2distribution}) are the most general type of 
Markovian quantum channels, describing classically correlated quantum errors. 
Here, the correlations are classical, because we work in the group 
${\rm SU}(2)\otimes {\rm SU}(2)$ -- for describing genuine quantum 
correlations, we would need to work in the group ${\rm SU}(4)$. 

In the following, we consider the simplest non-trivial case, correlated 
isotropic errors in both qubits. This model is able to interpolate 
between the case of independent single-qubit errors and identical errors 
(of different magnitude) applied to both qubits. In terms of the 
parametrization given in Eq.~(\ref{ME2:NormParam}) the model is characterized 
by setting $\vec b_1 = \vec b_2 = 0$ and 
\begin{equation}
\bm A_1 = a_1\, \One\; , \qquad \bm A_2 = a_2\, \One\; , \qquad 
 \bm F = \rho\, \sqrt{a_1\, a_2}\, \One\; ,
\label{TC:Lamc2}\end{equation}
where $a_1, a_2 > 0$ and $-1 \le \rho\le 1$. In the case of independent errors
(i.e. no correlations) $\rho = 0$, the model is equivalent to independent 
Pauli errors with $a_1 = -\, \ln(1 - 4p)$ and $a_2 = -\, \ln(1 - 4q)$, as can 
be seen from Eq.~(\ref{parametrosPauli}). For $\rho = 1$ the unitary errors 
applied to both qubits are maximally correlated. Note that $\rho = -1$ is quite 
peculiar: due to the lack of commutativity, it does not simply describe 
maximally anti-correlated errors.

For $0 < \rho < 1$ the present quantum channel, let us denote it as 
$\Lambda_{\rm c2}(\rho)$, interpolates between independent and maximally 
correlated quantum errors, just as the model based on Pauli errors, 
$\Lambda_{\rm cP}(m)$, introduced above. However, $\Lambda_{\rm c2}(\rho)$ is 
Lindblad divisible and hence Markovian by construction, even for intermediate
values of $\rho$.

Due to the simplicity of the model in question, the corresponding quantum 
channel can be expressed in closed form. In the notation of Sec.~\ref{ME2}, 
\begin{equation}
\mathcal{R}^{(2)}_{\rm c2} = \exp\big (\mathcal{L}^{(2)}\big )
 = 1\oplus \mathcal{R}_{\rm P}(p)\oplus \mathcal{R}_{\rm P}(q)
   \oplus \bm W_{\rm c2}\; ,
\end{equation}
where
\begin{equation}
\bm W_{\rm c2}= \begin{pmatrix}
	E_1&0&0&0&E_2&0&0&0&E_2\\
	0&E_3&0&E_4&0&0&0&0&0\\
	0&0&E_3&0&0&0&E_4&0&0\\
	0&E_4&0&E_3&0&0&0&0&0\\
	E_2&0&0&0&E_1&0&0&0&E_2\\
	0&0&0&0&0&E_3&0&E_4&0\\
	0&0&E_4&0&0&0&E_3&0&0\\
	0&0&0&0&0&E_4&0&E_3&0\\
	E_2&0&0&0&E_2&0&0&0&E_1\\
\end{pmatrix}\; ,
\label{TC:Wc2}\end{equation}
with coefficients
\begin{align}
E_1 &= E_0\; \frac{e^{2a_{12}} + 2\, e^{- a_{12}}}{3}\; , \quad
E_2 = E_0\; \frac{e^{2a_{12}} - e^{-a_{12}}}{3}\; , \notag\\
E_3 &= E_0\; \frac{e^{a_{12}} + e^{-a_{12}}}{2}\; , \quad
E_4 = E_0\; \frac{e^{-a_{12}} - e^{a_{12}}}{2}\; ,
\label{coeficientesE} \end{align}
where $E_0 = e^{-a_1 - a_2}$ and $a_{12} = \rho\, \sqrt{a_1 a_2}$.

In the case of independent errors, we find indeed that 
$\Lambda_{\rm cP}(0) = \Lambda_{\rm c2}(0)$. However, in general $m$ and $\rho$
cannot be related to one another such that 
$\Lambda_{\rm cP}(m) = \Lambda_{\rm c2}(\rho)$, not even for equal single qubit
errors, $p=q$ ($a_1 = a_2$). Interestingly, even for $m=\rho=1$ (maximal 
correlations) the two error channels are different. This is evident, when
comparing Eq.~(\ref{TI:WcP}) to Eq.~(\ref{TC:Wc2}).

Note that for $p_1 (p_2) =1/4$, we obtain a depolarizing channel 
which sends any (mixed) state $\varrho$ to the uniform mixture. In this case, 
the corresponding parameter $a_1$ ($a_2$) for the normal channel tends to 
infinity. This means that the normal distribution in SU$(2)$ tends to the 
uniform distribution (Haar measure) of the group.

\section{\label{sA} Application: Entanglement distillation}

In this section we consider the problem of 
entanglement distillation, where we expect to find significant differences 
when modeling the transmission errors either with $\Lambda_{\rm cP}(m)$ 
(correlated Pauli errors) or $\Lambda_{\rm c2}(\rho)$ (normal correlated 
errors). As a starting point, we consider the protocol introduced in 
Refs.~\cite{Ben96,BenMixEntangle}, where the sender (Alice) has two Bell pairs 
in the state
\begin{equation}
\ket{\Phi^+}=\frac{\ket{00}+\ket{11}}{\sqrt{2}}\; ,
\end{equation}
she wants to share with a receiver (Bob). For that she sends one qubit of each 
Bell pair through a noisy quantum channel to Bob. At this point, we assume that 
the qubits are sent together or one after the other through a noisy channel, 
where they pick up correlated errors.

It will be convenient to view this basic distillation protocol as a map from
the space of two-qubit quantum channels to the space of density matrices as
follows:
\begin{equation}
D \quad :\quad \Lambda_{\rm E2} \to \varrho = D\big ( \Lambda_{\rm E2}\big ) 
   \in\mathcal{S}(\mathbb{C}^4)\; ,
\end{equation}
where $\Lambda_{\rm E2}$ is the two-qubit error map. 

\begin{figure}
\includegraphics[width=\linewidth]{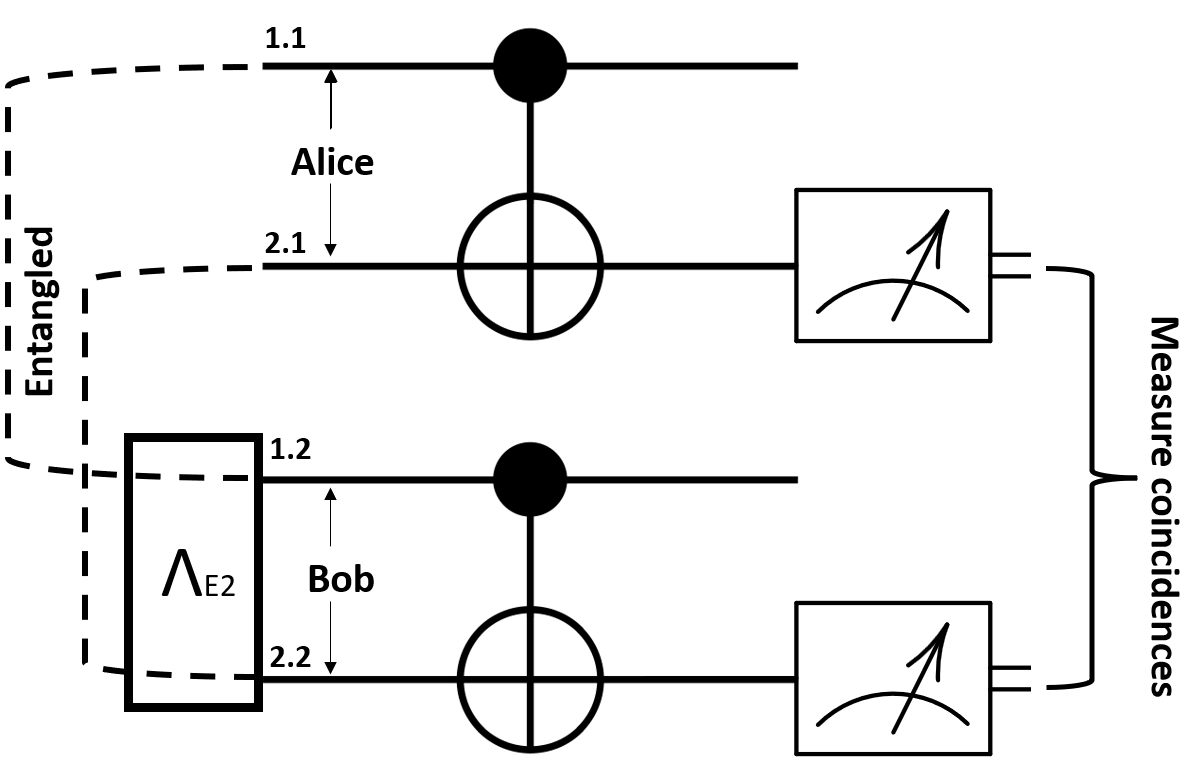}
\caption{The basic distillation process according to 
Refs.~\cite{Ben96,BenMixEntangle}. The horizontal lines represent the qubits
involved, the dotted lines indicate entanglement. The error map 
$\Lambda_{\rm E2}$ is shown in a box. The qubit pairs $1.k$ and $2.k$ start out 
in Bell states $\ket{\Phi^+_k}$. The pair $k.1$ ($k.2$) are in the possession of 
Alice (Bob). }
\label{fig:DiagramaDestilacion}\end{figure}

The distillation protocol $D$ is shown in Fig.~\ref{fig:DiagramaDestilacion}. 
Right after the transmission of the two qubits, Alice and Bob apply control-not 
gates to the two qubits in their possession, and measure $\sigma_z$ in one of 
those as shown in the diagram. They communicate their measurement results to 
each other. If their results coincide, they use the remaining qubits as an 
improved Bell pair, otherwise they discard the qubits and start anew. 

\begin{figure}
\includegraphics[width=\linewidth]{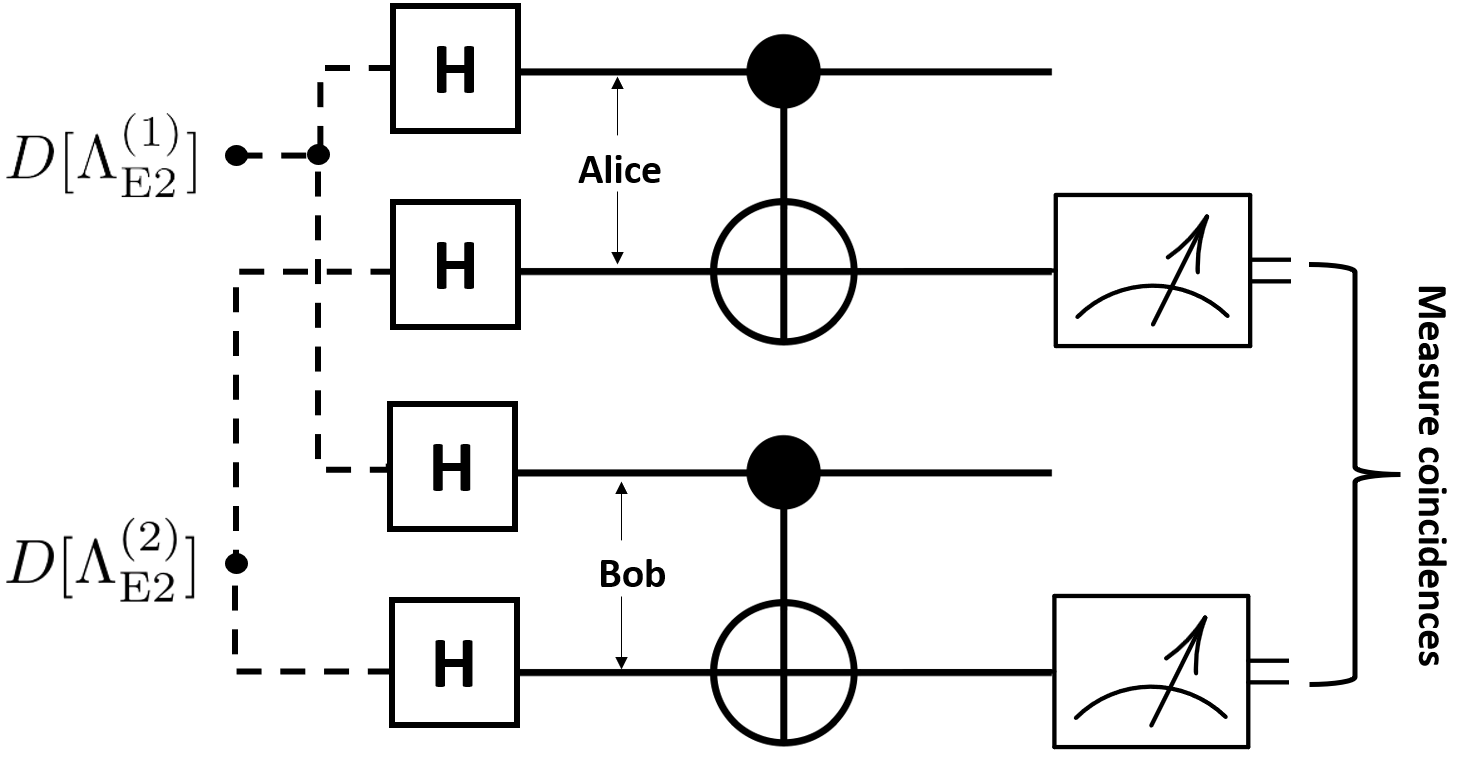}
\caption{Post-processing diagram after the application of $D^{\otimes 2}$. The 
horizontal lines represent the qubits involved, the dashed lines indicate the
approximate Bell states, after the initial distillation. }
\label{fig:PostProcess}\end{figure}

This protocol can fix errors of the type $\sigma_x$ or $\sigma_y$ but not 
$\sigma_z$~\cite{krastanov_albert_jiang_2019}. In order to correct arbitrary 
single-qubit errors, we apply the distillation protocol $D$ twice. This 
requires four Bell pairs at the beginning and yields two independent 
approximate Bell pairs as a result:
\begin{equation}
D^{\otimes 2}\; :\; \Lambda_{\rm E2}^{(1)}, \Lambda_{\rm E2}^{(2)} \to
 D\big (\Lambda_{\rm E2}^{(1)}\big )\otimes D\big (\Lambda_{\rm E2}^{(2)}\big )
 \in \mathcal{S}(\mathbb{C}^4)^{\otimes 2}\; ,
\end{equation}
where we allow that the error channel applied in the first distillation 
protocol is different from the one applied in the second. Afterwards, we apply 
a post-processing protocol $\Lambda_{\rm post}$ to the two approximate Bell 
pairs, such that the complete distillation protocol may be expressed as
\begin{equation}
D_{\rm u}(\Lambda_{\rm E2}) = \Lambda_{\rm post} \circ 
   D\big ( \Lambda_{\rm E2}\big )^{\otimes 2} \; .
\end{equation}
This post-processing protocol is shown in Fig.~\ref{fig:PostProcess}. It takes 
the two approximate Bell states resulting from the application of 
$D^{\otimes 2}$ and first applies Hadamard gates to all the qubits. Then it 
applies the same process as in $D$, which results in a single improved Bell
state at the end.

The protocol $D_{\rm u}$ can fix errors which include the three types 
$\sigma_x$, $\sigma_y$ and $\sigma_z$ at the cost of using four Bell states. To
check the effectivity of the protocol $D_{\rm u}$, consider that Alice uses the 
noisy channel to share only one Bell pair with Bob. In this case, both error
models, $\Lambda_{\rm cP}(m)$ and $\Lambda_{\rm c2}(\rho)$, reduce to a single 
qubit isotropic Pauli channel. The fidelity of the transmitted Bell pair is
\begin{align}
F_n &= \bra{\Phi^+} \left(\One \otimes \Lambda_{\rm P} \right)\big [ \,
   \ket{\Phi^+}\bra{\Phi^+}\, \big ]\; \ket{\Phi^+} \notag \\
    &= 1 -p_x -p_y -p_z.
\label{sA:noDistFid}\end{align}
If instead, Alice and Bob implement the distillation protocol $D_{\rm u}$, the
fidelity will improve. A simple result can be obtained for the same Pauli
errors in the absence of correlations. In this case, the fidelity becomes
\begin{align}
F_{\rm u} &= \bra{\Phi^+}\; D_{\rm u}\big (\, \Lambda_{\rm P}^{\otimes 2}\, 
   \big )\; \ket{\Phi^+} \notag \\
 &\approx 1 -2p_x^2 -2p_y^2 -4p_z^2\; , 
\label{sA:FidwDist}\end{align}
 when approximating the result up to second order in the probabilities 
$p_x, p_y, p_z$.

\begin{figure}
\includegraphics[width=\linewidth]{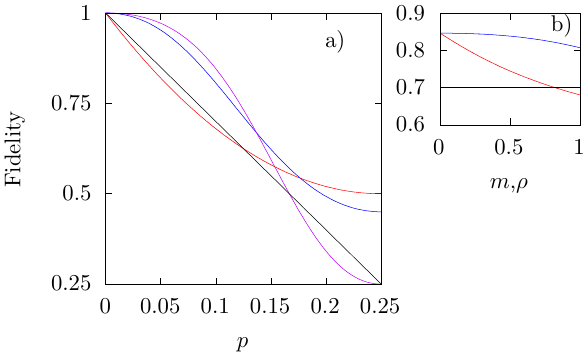}
\caption{Fidelity of the distillation protocol $D_{\rm u}$ for correlated Pauli 
errors $\Lambda_{\rm cP}(m)$ (red solid lines) and normal errors, 
$\Lambda_{\rm c2}(\rho)$ (blue solid lines).
Panel (a), fidelity as a function of $p$. Without distillation (black solid 
line); with distillation for uncorrelated errors $m=\rho=0$ (purple solid 
line), and for maximally correlated errors $m=\rho=1$.
Panel (b), fidelity as a function of $m,\rho$ for a fixed error probability
$p= 0.1$.}
\label{fig:fidel}\end{figure}

In the following, we consider the effect of correlations on the efficiency of 
the distillation process. In doing so, we compare correlated Pauli errors 
($\Lambda_{\rm cP}(m)$) with correlated normal errors 
($\Lambda_{\rm c2}(\rho)$). Figure~\ref{fig:fidel} shows the fidelity for the 
different cases, as a function of the error probability $p$ ($p=q$) in panel 
(a) and as a function of the degree of correlations in panel (b). 

In both panels, we show the case without distillation with a black solid line.
This case is described by Eq.~(\ref{sA:noDistFid}). Note that for $p=1/4$, the 
error channel becomes the completely depolarizing channel, which leaves the 
pair of qubits shared by Alice and Bob in the uniformly mixed state. Therefore
$F_n(p) = 1 - 3p$, and $F_n(1/4) = 1/4$.

Then we consider the fidelity $F_u$ under the distillation protocol $D_{\rm u}$
in the case of uncorrelated errors. Since 
$\Lambda_{\rm cP}(0) = \Lambda_{\rm c2}(0)$, both cases yield the same result,
shown by a single red solid line. As anticipated in Eq.~(\ref{sA:FidwDist}),
we see the gain in fidelity for small error probabilities, where the fidelity 
decay is quadratic in $p$.

Finally, let us discuss the case of maximal correlations. For the correlated
Pauli channel $\Lambda_{\rm cP}(1)$, this means that in 
Fig.~\ref{fig:DiagramaDestilacion} always the same Pauli errors are applied to 
qubits $1.k$ and $2.k$. For $p=1/4$, $D_{\rm u}$ leaves the two qubits in the 
state
\begin{equation}
D_{\rm u}\big ( \Lambda_{\rm cP}(1)\big ) = \frac{1}{2}\begin{pmatrix} 
   1 & 0 & 0 & 0\\ 0 & 0 & 0 & 0 \\ 0 & 0 & 0 & 0\\ 0 & 0 & 0 & 1\end{pmatrix}
\; .
\end{equation}
This state contains only classical correlations, and it results in 
$F_{\rm u} = 1/2$. In Fig.~\ref{fig:fidel}, panel (a), the result for the 
correlated Pauli channel is shown by a purple solid line. It ends indeed at 
$F_{\rm u}(1/4) = 1/2$. Note that for small values of $p$, the distillation
protocol cannot provide any improvement which is in line with our intuition.
Importantly, the situation is very different for the correlated normal channel,
$\Lambda_{\rm c2}(1)$. Here, small values of $p$ mean that random errors
are realized by unitaries which are close to the identity. In that case, even
for maximally correlated errors, the distillation protocol is 
almost as effective as in the uncorrelated case.

\begin{figure}
\includegraphics[width=\linewidth]{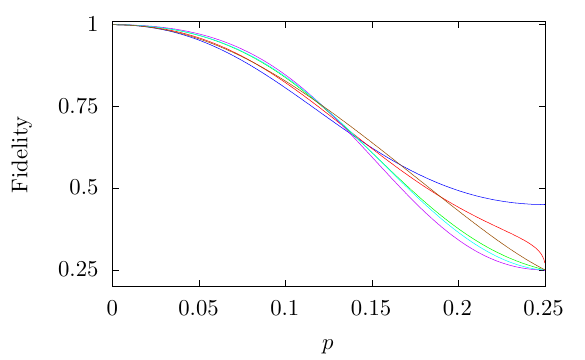}
\caption{Fidelity of $D_{\rm u}$ as a function $p$ for normal correlated 
errors, $\Lambda_{\rm c2}(m)$, varying $\rho$: blue ($\rho = 1$), red 
($0.8$), green ($0.5$), purple ($0$), cyan ($-0.50$), and brown 
($-1$). The gray dashed line is the undistilled case. }
\label{fig:fidelLambda2}\end{figure}

Figure~\ref{fig:fidelLambda2} shows the fidelity of the distillation protocol
$D_{\rm u}$ in the case of correlated normal errors, $\Lambda_{\rm c2}(\rho)$
as a function of the error probability $p$, for different values of $\rho$.
We can see that for small values of $p$ the protocol is always very effective,
and different values of the correlation coefficient $\rho$ yield very similar
results. This changes only at larger values $p \lesssim 1/4$. In this
figure we include two cases with anti-correlated errors $\rho < 0$, which show
that there is no symmetry between positive and negative correlations. For small 
errors, the correlations decrease the effectivity of the distillation, and the
highest fidelity is achieved for $\rho=0$. 

\section{\label{sC} Conclusions}

In this paper, we define normal distributions for compact Lie groups, based on 
the concept of infinite divisibility and derive explicit expressions for the 
associated normal quantum channels. 
This provides a general framework for describing errors in quantum processing
devices analogous to normally distributed errors in classical systems. 
The divisibility of the normal distributions allows us to view them as the 
result of a diffusive random walk in the group manifold. In the space of 
quantum channels, this gives rise to a differentiable quantum process, 
described by a Lindblad master equation.

We then focus on the groups ${\rm SU}(2)$ and ${\rm SU}(2)^{\otimes 2}$, which 
are particularly relevant for quantum information processing with qubits. 
In the case of ${\rm SU}(2)$ (single qubit case), we first show that the normal
channels are essentially equivalent to Pauli channels, which have always been
fundamental to modeling errors in quantum information science. Then, we 
demonstrate that different normal distributions may be associated to the same 
normal quantum channel. Intuitively, this can be understood in terms of the 
diffusive random walk picture. In this picture, different random walks describe 
different quantum processes, which nonetheless may arrive at the same quantum 
channel at the correct time.

In the case of ${\rm SU}(2)^{\otimes 2}$ (two qubit case), the normal 
distributions provide a flexible and intuitive framework for describing 
correlated quantum errors in the implementation of two-qubit quantum gates, or 
in the transmission of pairs of qubits across a noisy quantum channel. Choosing 
here ${\rm SU}(2)^{\otimes 2}$ over ${\rm SU}(4)$, amounts to restricting 
ourselves to single qubit errors with classical correlations. When comparing to
other models of correlated errors (e.g. models based on Pauli errors), our 
model has the advantage to be always Markovian, by definition. 

For illustration, we consider entanglement distillation in the presence of 
correlated isotropic single-qubit errors. In this example, we found important
differences between the model based on normal channels and the one based on
Pauli channels. For a small error rate, normal channels apply small random 
rotations to the qubits. Then, even in the presence of correlations, the 
distillation process remains very effective.

Our results may prove helpful in the tomography of imperfect quantum gates, in
particular two-qubit gates which often dominate the overall error in current
quantum computer platforms. They also help to give more accurate estimates of
the effect of such errors in different quantum protocols. 

In the case of classical devices, the method of error propagation is based on
the assumption of normally distributed errors. In this way, it provides a 
simple way to calculate the overall error in a given circuit of error prone 
operations. It may be possible to develop an analogous method for calculating 
accumulated errors in quantum circuits.

\acknowledgements

We are grateful to F. Leyvraz for very helpful discussions. This work has been 
financially supported by the Conacyt project ``Ciencias de Frontera 2019'', 
10872.

\appendix

\section*{Choi-matrix representation for single qubit normal channels} 

Here, we evaluate the Choi-matrix elements as defined in 
Eq.~(\ref{SC:defChoiDist}) writing the products
$D^{(1/2)}_{mm'}(g)\, D^{(1/2)}_{kk'}(g)^*$, as linear combinations of 
different Wigner D-matrices. For this, we take advantage of the general
structure of the Choi-matrix, to reduce the number of independent matrix 
elements which must be determined. In what follows we use the short-hand 
notation $\Lambda[|i\rangle\langle j|]_{kl} = \Lambda[i,j]_{kl}$. Then,
\begin{equation}
\mathcal{C}[f] = \begin{pmatrix}
 \Lambda[0,0]_{00} & \Lambda[0,0]_{01} & \Lambda[0,1]_{00} &
 \Lambda[0,1]_{01} \\
 \Lambda[0,0]_{10} & \Lambda[0,0]_{11} & \Lambda[0,1]_{10} &
 \Lambda[0,1]_{11} \\
 \Lambda[1,0]_{00} & \Lambda[1,0]_{01} & \Lambda[1,1]_{00} &
 \Lambda[1,1]_{01} \\
 \Lambda[1,0]_{10} & \Lambda[1,0]_{11} & \Lambda[1,1]_{10} &
 \Lambda[1,1]_{11} \end{pmatrix}
\end{equation}
Any quantum channel must preserve the Hermiticity -- this implies that the 
Choi-matrix itself must be Hermitian. Any quantum channel must preserve the
trace -- this implies that the diagonal elements of the diagonal blocks must 
sum to one, while the diagonal elements of the non-diagonal blocks must sum to 
zero. This leaves the following independent matrix elements to be determined:
\begin{equation}
\mathcal{C}[f] = \begin{pmatrix}
 \Lambda[0,0]_{00} & \ast & \ast & \ast \\
 \Lambda[0,0]_{10} & \ast & \ast & \ast \\
 \Lambda[1,0]_{00} & \Lambda[1,0]_{01} & \ast & \ast \\
 \Lambda[1,0]_{10} & \ast & \Lambda[1,1]_{10} & \Lambda[1,1]_{11} 
\end{pmatrix}\; .
\end{equation}
In addition, we see from Eq.~(\ref{generalnormalmap}) that any normal 
quantum channel must be unital, i. e. $\Lambda[\One] = \One$. Therefore, it 
must hold that $\Lambda[1,1]_{11} = \Lambda[0,0]_{00}$ and
$\Lambda[1,1]_{10} = - \Lambda[0,0]_{10}$. This leaves us with five independent
Choi-matrix elements to be determined.

\paragraph{Using the Peter-Weyl orthogonal basis.}
According to the Peter-Weyl theorem, the Wigner D-matrices form an orthogonal
basis for functions $\psi(g) \in \mathcal{L}^2({\rm SU}(2))$. The fact that 
they are not orthonormal is simply due to the different customary 
normalization~\cite{Sakurai94}. This implies,
\begin{align}
C^{(s)} &= \frac{1}{2s+1}\int dg\; \psi(g)\, D^{(s)}(g) \; , \notag\\
\Leftrightarrow\quad \psi(g) &= \sum_{s\in\mathbb{Z}_0^+/2} 
   {\rm tr}( C^{(s)}\, D^{(s)}(g)^\dagger) \; ,
\end{align}
which apart from the particular normalization is in line with 
Eq.~(\ref{aNH:PWONB}). With this it is an easy task to express the products
$D^{(1/2)}_{mm'}(g)\, D^{(1/2)}_{kk'}(g)^*$, as the following linear
combinations:
\begin{align}
&\begin{pmatrix}
   \Lambda[0,0]_{00} \\ 
   \Lambda[0,0]_{10} \\ 
   \Lambda[1,0]_{00} \\ 
   \Lambda[1,0]_{10} \\ 
   \Lambda[1,0]_{01} \end{pmatrix} \; :\;
\begin{pmatrix}
   D^{(1/2)}_{00}\, D^{(1/2)}_{00}{}^* \\ 
   D^{(1/2)}_{10}\, D^{(1/2)}_{00}{}^* \\ 
   D^{(1/2)}_{01}\, D^{(1/2)}_{00}{}^* \\ 
   D^{(1/2)}_{11}\, D^{(1/2)}_{00}{}^* \\ 
   D^{(1/2)}_{01}\, D^{(1/2)}_{10}{}^* \end{pmatrix} 
 = \begin{pmatrix} \frac{1}{2} + \frac{1}{2}\, D^{(1)}_{00}\\
     \frac{1}{\sqrt{2}}\, D^{(1)}_{-1,0}\\
     \frac{1}{\sqrt{2}}\, D^{(1)}_{0,-1}\\
     D^{(1)}_{-1,-1}\\
     -\, D^{(1)}_{1,-1}\\
   \end{pmatrix}\; ,
\label{AppA:Res}\end{align}
where we omitted the arguments ``$(g)$'' for brevity. Note that we used the 
fact that $D^{0}_{0,0} = 1$. 

%\bibliography{/home/gorin/Documentos/Bib/JabRef}
\bibliography{./JabRef}

\end{document}